\documentclass[12pt,preprint]{aastex}
\usepackage{graphicx}
\begin{document}
\title{Convective Instability of a Relativistic Ejecta Decelerated by a Surrounding Medium: 
An Origin of Magnetic Fields in GRBs?}
\author{Amir Levinson\altaffilmark{1}}
\altaffiltext{1}{Raymond and Beverly Sackler School of Physics \& Astronomy, Tel Aviv University,
Tel Aviv 69978, Israel; Levinson@wise.tau.ac.il}


\begin{abstract}
Global linear stability analysis of a self-similar solution describing a relativistic shell decelerated
by an ambient medium is performed.  The system is shown to be subject to the convective Rayleigh-Taylor
instability, with a rapid growth of eigenmodes having angular scale much smaller than the causality scale.
The growth rate appears to be largest at the interface separating the shocked ejecta and shocked ambient gas.
The disturbances produced at the contact interface propagate in the shocked media and cause nonlinear oscillations
of the forward and reverse shock fronts.  It is speculated that such oscillations may affect the emission 
from the shocked ejecta in the early afterglow phase of GRBs, and may be the origin of the magnetic field in the
shocked circum-burst medium.   
\end{abstract}
\keywords{gamma rays: bursts, instabilities, magnetic fields, hydrodynamics, relativity, turbulence}
\section{Introduction}
Following shock breakout and a rapid acceleration phase, the relativistic ejecta accumulated in
the explosion starts decelerating, owing to its interaction with the surrounding medium.  
At early times a double shock structure forms, consisting of a forward shock that propagates
in the ambient medium, a reverse shock crossing the ejecta and a contact interface separating
the shocked ejecta and the shocked ambient medium.  In the fireball scenario commonly adopted,
the naive expectation has been that the crossing of the reverse shock should produce an observable 
optical flash.  Despite considerable observational efforts, such flashes have not been detected yet.
The afterglow emission, which is produced behind the forward shock, seem to indicate strong
amplification of magnetic fields in the post shock region, by some yet unknown mechanism.  Moreover,
the lightcurve of the afterglow emission deviates at early time from that predicted by a simple blast 
wave model. 

A question of considerable interest is the stability of the double shock system.  Hydrodynamic instabilities
may lead to strong distortions of the system that may generate turbulence, amplify magnetic fields,
and affect the emission processes in the afterglow phase.  Such effects have been studied in the 
non-relativistic case in connection with young supernovae remnants (SNRs).
In fact, the idea that the Rayleigh-Taylor (R-T) instability may play an important role in 
the deceleration of a non-relativistic ejecta dates back to Gull (1973), who 
performed 1D simulations of young SNRs that incorporate a simple model of convection.
Chevalier et al. (1992) performed a global linear stability analysis of a 
self-similar solution describing the interaction of a nonrelativistic ejecta
with an ambient medium and found that it is subject to a convective instability.
They analyzed self-similar perturbations and showed that the flow is unstable
for modes having angular scale smaller than some critical value.
The convective growth rate was found to be largest at the contact discontinuity
surface and to increase with increasing $l$ number of the eigenmodes.  They
also performed 2D hydrodynamical simulations that verified the linear results and 
enabled them to study the nonlinear evolution of the instability.  The simulation
exhibits rapid growth of fingers from the contact interface that saturates, in the nonlinear state,
by the Kelvin-Helmholtz instability.  Strong distortion of the contact and the reverse shock was observed with little
effect on the forward shock.  Jun \& Norman (1996) performed 2 and 3D 
MHD simulations of the instability to study the evolution of magnetic fields in the convection
zone.  They confirmed the rapid growth of small scale structure reported by Chevalier et al. (1992), and in addition 
found strong amplification of ambient magnetic fields in the turbulent flow around R-T fingers. On average, the magnetic field energy
density reaches about 0.5\% of the energy density of the turbulence, but it could well be 
that the magnetic field amplification was limited by numerical resolution in their simulations.
The simulations of Chevalier et al. (1992) and Jun \& Norman (1996) support earlier ideas, that
the clumpy shell structure observed in young (pre-Sedov stage) SNRs such as Tycho, Kepler and Cas A 
is due to the R-T and K-H instability.  

In this paper we extend the linear stability analysis of Chevalier et al. (1992) into the relativistic regime.
We find that denser ejecta sweeping a lighter ambient gas are subject to the R-T instability also in the relativistic case.
The stability of a double-shock system has been investigated by Xiaohu et al. (2002) using the thin shell approximation.  However,
this study is limited to large scale modes and neglects pressure gradients and, therefore, 
excludes the convective instability. 
Gruzinov (2000) performed a linear stability analysis of a Blandford-McKee (BMK) blast wave solution, and  
found that the BMK solution is stable but non-universal,
in the sense that some modes decay very slowly as the system evolves.  Furthermore, the onset of oscillations
of an eigenmode of order $l$ has been seen in the simulation once the Lorentz factor evolved to 
$\Gamma <l$.  The conclusion drawn based on Gruzinov's findings 
is that distortion of the shock front at early times may 
cause significant oscillations during a large portion of its evolution.  If the amplitude 
of these oscillations is sufficiently large, and if the same behavior holds in 
the nonlinear regime then this can lead to generation of vorticity in the post shock region 
(Milosavljevic et al. 2007; Goodman \& MacFadyen 2007), and the consequent amplification of magnetic 
fields, as demonstrated recently by Zhang et al. (2009).

\section{Analysis}

We consider the interaction of a cold unmagnetized shell with a cold ambient medium 
having a density profile $\rho_i=br^{-k}$.  The structure 
formed at early stages  consists of a forward shock propagating in the ambient medium, a reverse shock propagating 
in the ejecta and a contact discontinuity separating the shocked ambient medium and 
the shocked ejecta.  The equations governing the dynamics of the flow on each side of the contact interface 
can be written in the form,
\begin{eqnarray}
\rho h\gamma^2\frac{d\ln\gamma}{dt}+\gamma^{2}\frac{dP}{dt}=\frac{\partial P}{\partial t},\label{momentum}\\
\frac{d}{dt}\ln\left(P/\rho^{\hat{\gamma}}\right)=0,\label{state}\\
\rho\gamma\frac{d}{dt}(h\gamma{\bf v}_T)+\nabla_T P=0,\label{tangent}
\end{eqnarray}
where $\gamma=u^0$ is the Lorentz factor of the fluid,
${\bf v}_T$ is the tangential component of the 3-velocity, which we express as 
${\bf v}=v_r\hat{r}+{\bf v}_T$, $\rho$, $P$ and $h$ are the proper density, pressure and specific enthalpy, 
$\hat{\gamma}$ is the adiabatic index, and 
\begin{equation}
\nabla_T\equiv \frac{1}{r}\frac{\partial}{\partial\theta}\hat{\theta}+\frac{1}{r\sin\theta}
\frac{\partial}{\partial\phi}\hat{\phi}.
\end{equation}
Equations (\ref{momentum})-(\ref{tangent}) admit self-similar solutions (Nakamura \& Shigeyama, 2006)
in cases of a freely expanding ejecta characterized by a velocity $v_e=r/t$ at time $t$ after the explosion, and  
a proper density profile
\begin{equation}
\rho_e=\frac{a}{t^3\gamma_e^n}, \label{den-ejecta}
\end{equation}
where $\gamma_e=1/\sqrt{1-v_e^2}$ is the corresponding Lorentz factor of the unshocked ejecta.
The Lorentz factors of the forward shock, reverse shock and the contact discontinuity surface, 
denoted by $\Gamma_{1}(t)$, $\Gamma_{2}(t)$ and $\Gamma_c(t)$, respectively, evolve with time 
as $\Gamma^2_2=At^{-m}$, $\Gamma^2_1=Bt^{-m}$, $\Gamma^2_c=Ct^{-m}$, with the constants
$A,B,C$ and $m$ determined by matching the solutions in the shocked ejecta and in the shocked ambient 
medium at the contact discontinuity.  In particular $m=(6-2k)/(n+2)$  (Nakamura \& Shigeyama, 2006).  
The velocity of the ejecta just upstream the reverse shock is $v_e(r=r_2)=
r_2(t)/t=1-1/[2(m+1)\Gamma_2^2]$, where $r_2(t)=\int{(1-1/2\Gamma_2^2) dt}$ is trajectory of the 
reverse shock, implying $\gamma_e^2=(m+1)\Gamma_2^2$.  Thus, the self-similar solution is applicable
only to situations where the ejecta is sufficiently dense, such that the reverse shock is non or at most
mildly relativistic.

We have carried out a global linear stability analysis of the self-similar solution outlined above.  The 
details will be presented elsewhere (Levinson, in preparation).  A preliminary account of the method and results
is presented below.

There are total of eight independent variables, four on each side of the contact: $P,\rho, \gamma, {\bf v}_T$.
The perturbations of these variables were expanded in spherical harmonics.  To be more concrete, a perturbed 
quantity $Q$  ($Q=P, \rho$, etc.) is expressed as $Q(t,\chi,\theta,\phi)=
Q_0(t,\chi)[1+\xi_Q(\chi,t)Y_{lm}(\theta,\phi)]$,
where $\chi=\{1+2(m+1)\Gamma_{1}^2\}(1-r/t)$ is the self-similarity parameter, and $Q_0$ denotes 
the unperturbed value.   The linearized equations on each side of the contact discontinuity 
were then obtained upon substitution of the perturbed quantities into Eqs. (\ref{momentum})-(\ref{tangent}).
Perturbations of the shock fronts and the contact discontinuity of the form
\begin{equation}
\delta r_j(t,\theta,\phi)=\frac{t\delta_j(t)}{\Gamma_j^2}Y_{lm}(\theta,\phi),\label{del_rs}
\end{equation}
where $j=1,2,c$ refers to the forward shock, reverse shock, and the contact discontinuity, respectively, 
were assumed.  The perturbed shock normals are then $n_{k\mu}=n_{k\mu}^{0}+\delta n_{k\mu}$ ($k=1,2$),
with $n_{k\mu}^{0}=(-\Gamma_{k}V_{k},\Gamma_{k},0)$ being the unperturbed normal and 
\begin{equation}
\delta n_{k\mu}=(-\Gamma^3_{k}\delta V_{k},\Gamma^3_{k} V_{k}\delta V_{k},
-\Gamma_{k}\nabla_T\delta r_{k}).\label{del_n}
\end{equation}
Here $V_{k}$ denotes the 3-velocity of the unperturbed shock front and $\delta V_k=d\delta r_k/dt$.
The linearized jump conditions at the forward and reverse shocks are given in terms of the 4-velocity $u^\mu$
and the energy-momentum tensor $T^{\mu\nu}$ as:
\begin{eqnarray}
[\rho u_0^{\mu}\delta n_{k\mu}+\Delta_k (\rho u^{\mu})n^0_{k\nu}]=0,\label{jmp1}\\
\left[T_0^{\mu\nu}\delta n_{k\nu}+\Delta_k T^{\mu\nu}n^0_{k\nu}\right]=0,\label{jmp2}
\end{eqnarray}
where $\Delta_k Q=\delta Q+(\partial Q_0/\partial r)\delta r_k$ denotes the Lagrange perturbation of 
the quantity $Q$ at the perturbed surface $k$.  The relations (\ref{jmp1}), (\ref{jmp2})
provide 6 boundary conditions for the perturbation equations, 3 on each side of the contact.
Two additional boundary conditions are obtained from pressure balance and the no flow condition,
viz., $v-dr_c/dt=0$, at the contact discontinuity.  

Unlike in the non-relativistic case, the boundary conditions at the shock fronts break self-similarity 
of the perturbations.  Specifically, it can be 
shown that at the forward shock surface the perturbation of the tangential velocity is related to the 
perturbations of the radial velocity and pressure through $\xi_T=3(\xi_P-\xi_R)/[2(m+2)\Gamma_1^2]$ for $l\ne0$, 
with a similar relation, though somewhat more involved, at the reverse shock front.  Thus, numerical simulations
of the perturbation equations is needed.  The only exception
is the spherical mode ($l=0$) for which a self-similar solution was obtained analytically.  For this 
solution $\delta Q\propto t^s$ with $s<0$.  It can be shown that one eigenmode of order $l=0$ is associated with 
linear time translation of the self-similar solution.  For this mode $s=-(m+1)$.  We found another eigenmode of order 
$l=0$ that decay somewhat slower.  The analytic solution for the $l=0$ mode
has been used both to test the code and as initial condition for the evolution of the higher order 
eigenmodes.  Numerical integration of the perturbation equations was performed
after transforming to the so called Riemann invariants.  We identified three variables that 
propagate from the forward shock inwards, three that propagate from the reverse shock outwards
and two that propagate from the contact, one inwards and one outwards, and have chosen the 
boundary conditions for the Riemann invariants accordingly.  It is generally found that eigenmodes having
an angular scale larger than the horizon scale, specifically $l(l+1)<\Gamma^2$, are stable.  Higher 
order modes are found to be unstable with a growth rate that increases with increasing $l$.  
An example is exhibited in Fig. 1.  The onset of oscillations followed by a rapid growth of the initial 
perturbations is clearly seen.   The distortion of the contact discontinuity surface becomes nonlinear
very early on.  The growth is algebraic in time $t$ with a growth rate of about $10$ in the example shown 
in Fig. 1; that is, $\delta Q\propto t^{10}$ for $Q=P,\rho, v$.  As seen, the reverse shock responds quickly to 
the distortion of the contact.  The forward shock, on the other hand, responds much later, at time when the 
instability near the contact already reached a nonlinear state.

\section{Discussion}
The stability analysis described above seem to indicate strong convective instability at early 
stages of the evolution of a dense ejecta as it sweeps a lighter ambient gas.  The growth rate
appears to be largest at the contact discontinuity and for higher order modes.  Disturbances at 
the interface separating the shocked ejecta and the shocked ambient medium propagate 
away from the contact discontinuity and cause nonlinear distortions of the shock fronts.  The reverse
shock responds quickly to the distortion of the contact.   Propagation of the signal to the forward shock is 
much slower.  At any rate, the instability near the contact becomes nonlinear well before the signal arrives at 
the forward shock, so full MHD simulations are needed to resolve the effect of the instability on the forward shock.
It is naively expected that the instability will be strongly suppressed in cases where the ejecta is highly 
magnetized and/or if the reverse shock is highly relativistic.  On the other hand, if the magnetic field 
strength in the unshocked ejecta is smaller than that required to suppress the instability but still much larger than 
that of the ambient medium, then at early stages mixing
of the magnetized ejecta with the shocked ambient gas via growth of R-T fingers can give rise to a 
strong amplification of the magnetic fields behind the forward shock.  Full simulations are required to quantify
the conditions under which the instability is effective. 

The nonlinear distortions of the contact and the shock fronts should generate turbulence
in the shocked fluids on both sides of the contact discontinuity.  At early stages this may strongly 
affect particle acceleration and the emission processes.  It is tempting to speculate that 
the lack of observed optical flashes, that are anticipated in the ``standard'' model, and the fact that the early 
afterglow emission observed in many sources is inconsistent with the prediction of the blast wave
model may be attributed to the instability discussed here.  In any case, it is clear that a careful 
analysis that takes account of this process is required to better understand the observational characteristics
of the emission during the early afterglow phase.

The stability analysis of the Blandford-McKee solution performed by Gruzinov (2000) suggests that 
it may be a very slow attractor.  Linear perturbations
of the forward shock in the B-M phase decay very slowly.  Whether this behavior continues also 
in the nonlinear regime is unclear yet.  If it does then it is anticipated that the growth of R-T fingers and, perhaps, 
nonlinear oscillations of the forward shock itself that are induced by the convective instability may be a source of vorticity
during a long portion of the evolution of the blast wave.  As demonstrated recently by Zhang et al. (2009) the induced 
turbulence can amplify weak magnetic fields. Their simulation seem to converge at a saturation level 
of $\epsilon_B\sim5\times10^{-3}$, weakly dependent on the initial magnetic field strength.  This process may 
provide an explanation for the origin of the strong magnetic fields inferred behind the collisionless shock 
in the afterglow phase.

Unfortunately, the linear analysis outlined above is restricted to a limited set of conditions under which the 
unperturbed self-similar solution of Nakamura \& Shigeyama (2006) is applicable.  
Full 3D MHD simulations are required to study this process in other situations, and to follow the  
evolution of the convective instability in the nonlinear state.  As illustrated above, high resolution 
simulations that can resolve angular scales $\Delta \theta<<1/\Gamma$ are required, posing a great numerical challenge.
We believe that our findings strongly motivate such efforts.

I thank A. Ditkowski for a technical help in the development of the code,
and M. Alloy, A. MacFadyen, E. Nakar and E. Waxman for enlightening discussions.
This work was supported by an ISF grant for the Israeli Center for High Energy Astrophysics,
and by the NORDITA program on Physics of relativistic flows.

\begin{figure}[h]
\centering
\includegraphics[width=16cm]{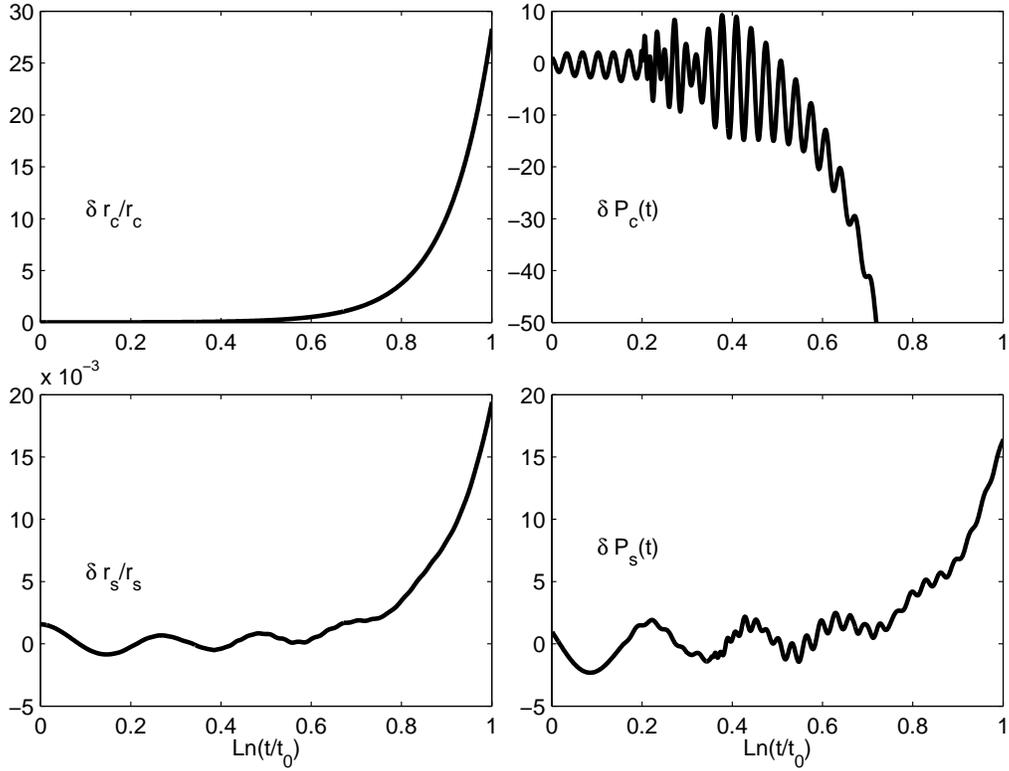}
\caption{ Time evolution of the perturbations for $n=1.1$, $k=2$ and $l(l+1)/\Gamma_1^2(t_0)=10^4$, 
here $\Gamma_1(t_0)$ is the 
initial Lorentz factor of the forward shock. Upper panels: perturbation of the contact discontinuity surface (left panel)
and relative pressure perturbation, $\delta P(t)/\delta P(t_0)$, of the shocked ambient medium at the contact (right panel). 
Bottom panels: perturbations of the reverse shock surface (left), and relative pressure perturbation at the reverse
shock front (right).  The initial perturbations of the contact discontinuity and the reverse shock surface in this example 
are $\delta r_c/r_c=6\times10^{-3}$ and $\delta r_s/r_s=10^{-3}$, respectively.}
\end{figure}


\begin{thebibliography}{99}
\bibitem{c92} Chevalier R.~A., Blondin, J.~M. \& Emmering R. 1992, ApJ, 392, 118
\bibitem{GM07} Goodman, J. \& MacFadyen, A. 2007,  arXiv:0706.1818
\bibitem{G00} Gruzinov, A. 2000, arXiv:astro-ph/0012364
\bibitem{Gu73} Gull, S.~F. 1973, MNRAS, 161, 47 
\bibitem{JN96} Jun, B-I. \& Norman, M.~L. 1996, ApJ, 465, 800
\bibitem{MN07} Milosavljevic, M., Nakar, E. \& Zhang, F. 2007,  arXiv:0708.1588
\bibitem{NS06} Nakamura, K. \& Shigeyama, T. 2006, ApJ, 645, 43
\bibitem{Xi02} Xiaohu, W., Loeb, A. \& Waxman, E. 2002, ApJ, 568, 830
\bibitem{ZM09} Zhang, W., MacFadyen, A. \& Wang, P. 2009, ApJ, 692, 40
\end{thebibliography}
\end{document}